\documentclass[twocolumn]{aastex62}

\usepackage{subfigure}
\usepackage{url}
\usepackage{hyperref}
\usepackage{epsfig}
\usepackage{graphicx}
\usepackage{sidecap}
\usepackage{hanging}
\usepackage{color}
\usepackage{multirow}
\usepackage{enumerate}
\usepackage{natbib}
\usepackage{amssymb}
\usepackage{amsmath}
\usepackage[bottom]{footmisc}

\newcommand{\sci}{Science}

\newcommand{\Cassini}{{\it Cassini}}
\newcommand{\Kepler}{{\it Kepler}}

\newcommand{\HST}{{\it HST}}
\newcommand{\JWST}{{\it JWST}}

\newcommand{\Spitzer}{{\it Spitzer}}

\providecommand{\e}[1]{\ensuremath{\, \times \, 10^{#1}}}

\received{2019 February 2}
\revised{2019 February 28}
\accepted{2019 February 28}

\submitjournal{ApJ Letters}

\shorttitle{Transit Recovery of Kepler-167e}
\shortauthors{Dalba \& Tamburo}

\begin{document}

\title{\textit{Spitzer} Detection of the Transiting Jupiter-analog Exoplanet Kepler-167e}

\correspondingauthor{Paul A. Dalba}
\email{pdalba@ucr.edu}


\author[0000-0002-4297-5506]{Paul A. Dalba}
\affiliation{Department of Earth \& Planetary Sciences, University of California Riverside, 900 University Avenue, Riverside, CA 92521, USA}

\author[0000-0003-2171-5083]{Patrick Tamburo}
\affiliation{Department of Astronomy, Boston University, 725 Commonwealth Avenue, Boston, MA 02215, USA}


\begin{abstract}

We acquired observations of a partial transit of Kepler-167e, a Jupiter-analog exoplanet on a 1071~day orbit, well beyond its water ice line, with the \textit{Spitzer Space Telescope}. The timing of the \textit{Spitzer} transit is consistent with the ephemeris measured from the two transits observed previously by the \textit{Kepler Space Telescope}. The \textit{Spitzer} observation rules out the existence of transit timing variations (TTVs) on the order of hours to days that are known to exist for other long-period exoplanets. Such TTVs render transit follow-up efforts intractable due to the substantial observing time required and the high risk of nondetection. For Kepler-167e, however, we are now able to predict future transit times through the anticipated era of the \textit{James Webb Space Telescope} with uncertainties of less than six minutes. We interpret the lack of TTVs as an indication that Kepler-167e either does not have an exterior massive companion or that the gravitational interactions with any companions are below our detection threshold. We also measure Kepler-167e's 3.6~$\mu$m transit depth and use exoplanet and solar system models to make predictions about its transmission spectrum. The transiting nature of Kepler-167e and its similarity to Jupiter make it a unique and exceptional target for follow-up atmospheric characterization. Kepler-167e falls into a truly rare category among transiting exoplanets, and with a precisely constrained transit ephemeris, it is poised to serve as a benchmark in comparative investigations between exoplanets and the solar system.

\end{abstract}



\section{Introduction} \label{sec:intro}

Jovian planets are excellent candidates for comparative investigations between the solar system and exoplanets. In the solar system, the dynamics of Jupiter and Saturn likely sculpted the system's final orbital architectures \citep[e.g.,][]{Tsiganis2005,Morbidelli2007a,Batygin2015}. The Jovian planets also likely influenced the properties of Mars \citep{Walsh2011} and the Earth, specifically related to the delivery of volatiles, the impact rates of minor bodies, and the development of life \citep[e.g.,][]{Zahnle1997,Morbidelli2000,Horner2010}. In exoplanetary systems, evidence exists for the substantial migration of Jovian exoplanets \citep[e.g.,][]{Lin1996}. A full understanding of the mechanisms dictating this migration will come with repercussions for the occurrence rates and architectures of planetary systems hosting giant planets. This is equally true for the solar system, where such migration appears not to have occurred \citep[e.g.,][]{Masset2001,Morbidelli2007b}.

Although the discovery of exoplanets has so far been dominated by the transit method, Jovian planets are more readily discovered by radial velocity (RV) surveys. Unfortunately, atmospheric characterization of RV exoplanets \citep[at least those that do not transit;][]{Dalba2019} awaits the next generation of direct imaging observatories \citep[e.g., the \textit{Wide Field Infrared Survey Telescope},][]{Spergel2015}.

Alternatively, follow-up characterization of Jovian-analog exoplanets (Jovian exoplanets with orbital properties or stellar irradiation resembling Jupiter or Saturn) is feasible via transit observations. \Cassini\ spacecraft observations demonstrated the amenability of Saturn-analog exoplanets to transmission spectroscopy \citep{Dalba2015}. Atmospheric refraction provides another means of atmospheric characterization and one that is specifically tailored to Jovian-analog exoplanets \citep{Dalba2017b}. Additionally, repeated transit observations offer an opportunity to detect and characterize exomoons \citep[e.g.,][]{Kipping2011,Teachey2018a,Teachey2018b}.

However, the low transit probability of a Jovian-analog exoplanet is a major obstacle to their characterization. Despite this crippling bias, the four-year baseline of the primary \Kepler\ mission \citep{Borucki2010} enabled the discovery of Kepler-167e, a 0.9~$R_{\rm Jupiter}$ exoplanet orbiting a K-dwarf host star once every $\sim$1071~days \citep{Kipping2016b}. Kepler-167e has been branded as a Jupiter-analog because of its size, its low eccentricity, and its Jupiter-like stellar insolation. \Kepler\ observed only two transits of Kepler-167e, which did not allow for the detection of transit timing variations (TTVs). Around 50\% of \Kepler's long-period transiting exoplanets and candidates exhibit TTVs of at least 2--40~hr \citep{Wang2015}. These variations are difficult to characterize with only a few observed transits and, for many cases, leave an insuperable uncertainty on the timing of future transits. 

Here, we present observations acquired by the \textit{Spitzer Space Telescope} of a partial transit of Kepler-167e. These observations confidently rule out the existence of TTVs that would have precluded characterization of this Jupiter-analog exoplanet with the \textit{Hubble Space Telescope} (\HST) and the upcoming \textit{James Webb Space Telescope} (\JWST). In Sections \ref{sec:spitzer_obs} and \ref{sec:spitzer_analysis}, we summarize the \Spitzer\ observations and data analysis. In Section \ref{sec:ephem}, we analyze the \Kepler\ transits in conjunction with the \Spitzer\ transit to constrain the magnitude of TTVs and to predict the transit times of Kepler-167e through 2030. Lastly, in Section \ref{sec:disc}, we discuss the immense potential of Kepler-167e as a target for follow-up characterization.


\section{{\it Spitzer} Observations}\label{sec:spitzer_obs}

The \textit{Spitzer Space Telescope} observed Kepler-167e for 10 consecutive hours beginning on 2018 December 14 at 14:31 Coordinated Universal Time (UTC) under program 14047 (PI: P. Dalba). The \Spitzer\ observations were designed as in \citet{Dalba2016}, to test the ephemeris of an exoplanet with only two previously observed transits for TTVs. The brightness of Kepler-167 ($K=11.832$) coupled with the anticipated transit depth ($\sim$1.6\%) meant that this test did not require the observation of a full transit. In the absence of TTVs, we expected the observation window to be (roughly) centered on transit egress. 

Observations were made using Channel 1 (3.6~$\mu$m) of the Infrared Array Camera \citep[IRAC,][]{Fazio2004} in accordance with the \Spitzer\ best practices for high-precision photometry.\footnote{See \url{https://irachpp.spitzer.caltech.edu/page/Obs\%20Planning}.} The systematic noise is known to be $\sim$4 times higher in Channel 1 than in Channel 2 \citep[e.g.,][]{Krick2016}. Still, we observed in Channel 1 where Kepler-167 is brighter and where opacity from CH$_4$ and higher-order hydrocarbons could have potentially affected the 3.6~$\mu$m transit depth \citep[e.g.,][]{Dalba2015}. In total, 2648 10.4~s exposures of Kepler-167 were acquired.


\section{{\it Spitzer} Data Analysis}\label{sec:spitzer_analysis}

IRAC observations suffer from systematic noise due to intrapixel variations in quantum efficiency (i.e., the ``pixel phase effect''). Here, we describe our methods to simultaneously account for this effect and fit the transit light curve of Kepler-167e.
 
To mitigate the pixel phase effect, we utilized Pixel-Level Decorrelation \citep[PLD,][]{Deming2015}, which has been extensively used to derive accurate transit and eclipse depths from \Spitzer\ data \citep[see][and references therein]{Tamburo2018}. For this observation, we modified this regression procedure to additionally choose the optimal grid of basis pixels. First, we created 25 different sets of basis pixel grids, starting with the single brightest pixel and including the next faintest pixel in subsequent grids as determined by the average flux captured over the time series. Then, for each combination of aperture radius and bin size, we performed a regression fit to the raw photometry for each of the 25 sets of pixels. The grid that optimized the Bayesian Information Criterion \citep{Schwarz1978} was chosen for that particular combination of aperture radius and bin size. Finally, we used the Allan variance criterion to choose the best global regression solution. 

\begin{figure}
    \centering
    \includegraphics[width=\columnwidth]{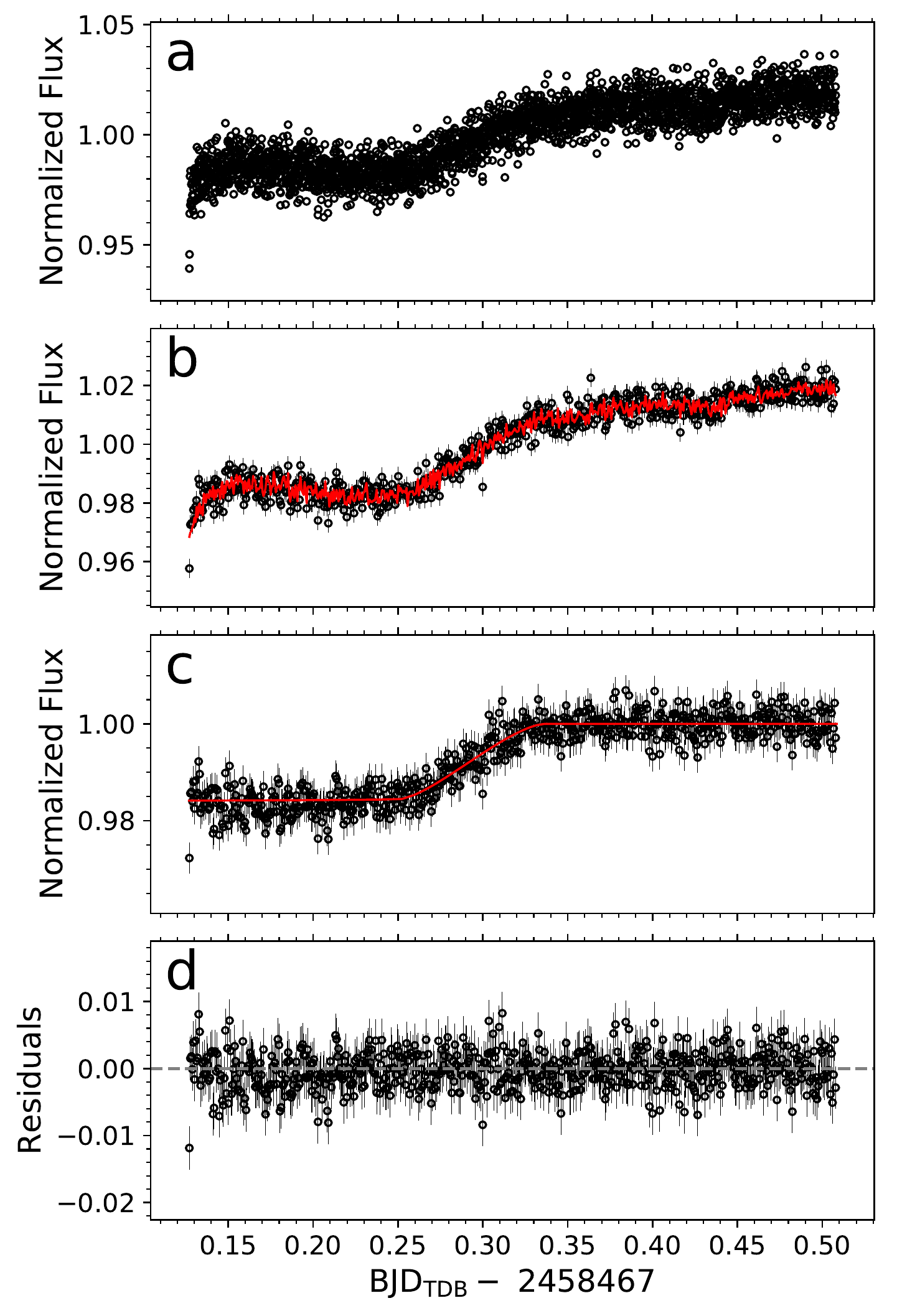}
    \caption{Panel (a): normalized raw \Spitzer\ photometry of Kepler-167. Panel (b): binned photometry (black points) and best PLD regression solution (red curve). Panel (c): photometry corrected for the pixel phase effect and the best-fit transit model. Panel (d): residuals between the data and model in Panel (c). The rms of the residuals is 0.279\%.}
    \label{fig:PLD}
\end{figure}

Our procedure selected a basis pixel grid consisting of the three brightest pixels (with coefficients ``Pix. 1,'' ``Pix. 2,'' and ``Pix. 3''). The time series intensities of these pixels were normalized following \citet{Deming2015} and then fit to the photometry along with a linear temporal ramp function (with slope $c_1$ and offset $c_0$) and a transit model. The optimal PLD solution used photometry extracted from a 1.6~pixel radius aperture, binned by a factor of four. The photometry throughout each stage of this analysis is shown in Figure \ref{fig:PLD}. 

We then used the Markov chain Monte Carlo (MCMC) formalism described by \citet{Deming2015} to fit a transit model to the \Spitzer\ data. We applied Gaussian priors to inclination ($i$) and the semi-major axis ($a$) scaled to the stellar radius ($R_{\star}$) using the parameters from \citet{Kipping2016b}. The quadratic limb darkening parameters ($u_1$ and $u_2$) were also given Gaussian priors centered at the values from \citet{Claret2011} that most closely match Kepler-167 and with width 0.2. Mid-transit time ($t_{\rm mid}$) and the planet radius ($R_p$) scaled to the stellar radius were given uniform priors spanning the entire physically allowed parameter space. Alternatively, the orbital period ($P$), eccentricity ($e$), and longitude of periastron ($\omega$) were locked to their median values from \citet{Kipping2016b}. Fitting the partial \Spitzer\ transit for these parameters would have yielded weak constraints and would have unrealistically inflated the uncertainties of the other parameters when marginalized over $P$, $e$, and $\omega$. 

\begin{figure}
    \centering
    \includegraphics[width=\columnwidth]{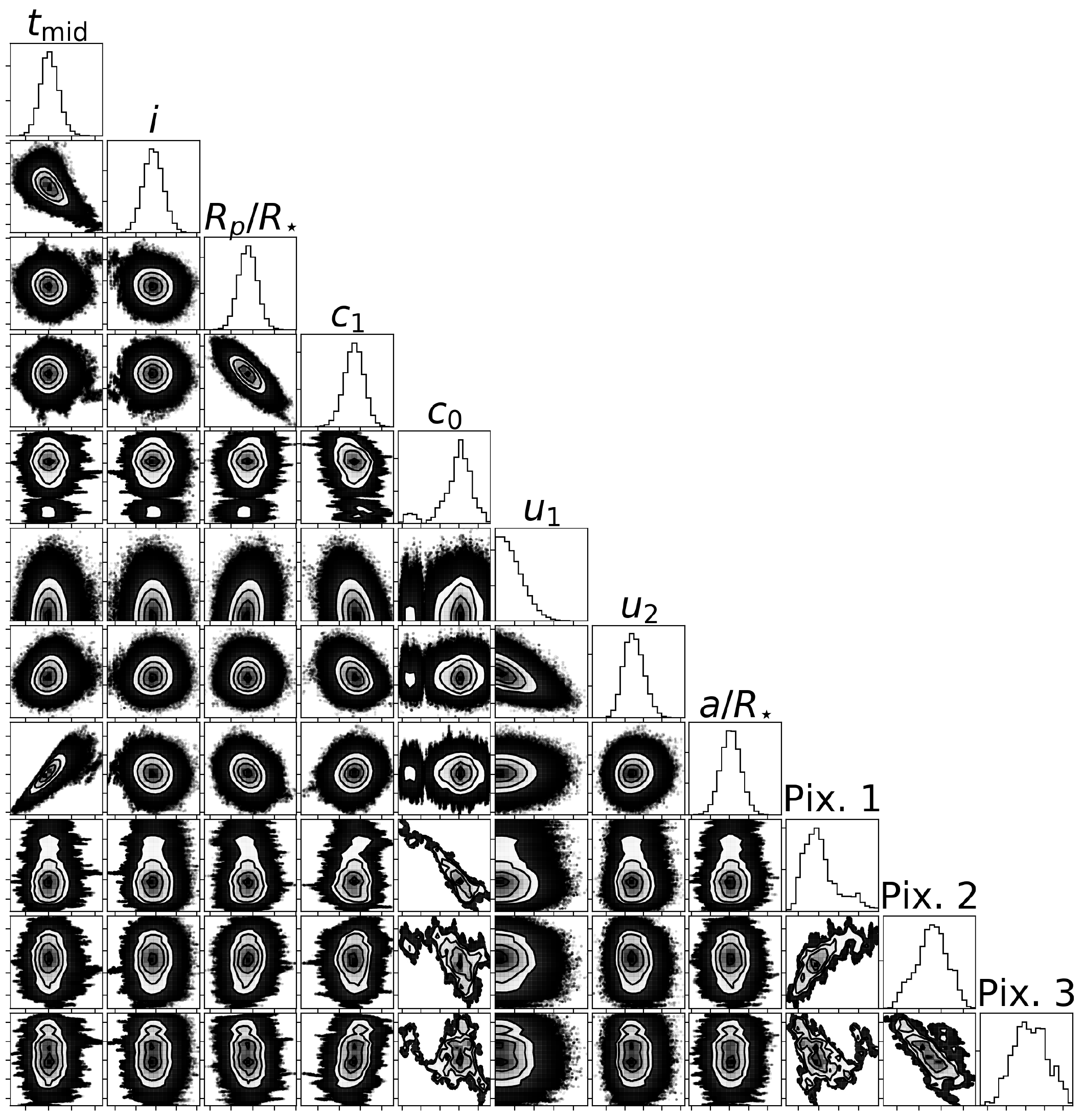}
    \caption{Posterior probability distributions for the parameters in the PLD fit to the \Spitzer\ data \citep[made with \texttt{corner},][]{ForemanMackey2016a}.}
    \label{fig:corner}
\end{figure}

We ran three MCMC chains of 1\e{6}~steps each, with 1\e{4}-step burn-in periods. This burn-in was chosen to adjust the step size for each parameter to converge on an acceptance rate of 35\%. We confirmed the convergence of the chains using the Gelman--Rubin statistic, which was at most 1.01 for the fitted physical parameters. Not all of the systematic parameters (i.e., $c_0$ and Pix. 1--3) achieved this level of convergence, as demonstrated by the posterior probability distributions in Figure \ref{fig:corner}. This is not unusual for PLD \citep[e.g., see Figure 11 of][]{Deming2015}. However, the physical parameters do not correlate with these systematic parameters, so the impact on the inferred physical parameter values is negligible. As expected, our constraints on $i$ and $a/R_{\star}$ are consistent with, but less precise than, those reported by \citet{Kipping2016b}. We measured the mid-transit time and planet--star radius ratio to be $t_{\rm mid}~=~2458466.985^{+0.016}_{-0.015}$~BJD$_{\rm TDB}$\footnote{BJD$_{\rm TDB}$ is Barycentric Julian Date (BJD) expressed as Barycentric Dynamical Time (TDB).} and $R_p/R_{\star}~=~0.1260^{+0.0035}_{-0.0035} $. These uncertainties are based on the 16th and 84th percentiles of the posterior probability distributions.  

To test the reliability of our basis pixel selection, and to ensure that the small basis set was not merely the result of the BIC's complexity penalty, we completed our entire \Spitzer\ data analysis again using the nine brightest pixels in PLD as the basis set. The physical (transit) parameters achieved the same convergence using this new model and their credible intervals were statistically indistinguishable from those shown in Figure \ref{fig:corner}. This result increases the confidence in the basis pixel selection procedure and in the resulting interpretation of the \Spitzer\ data.

\section{Confirmation of the Linear Ephemeris of Kepler-167\MakeLowercase{e}}\label{sec:ephem}

\begin{figure*}
    \centering
    \includegraphics[width=0.85\textwidth]{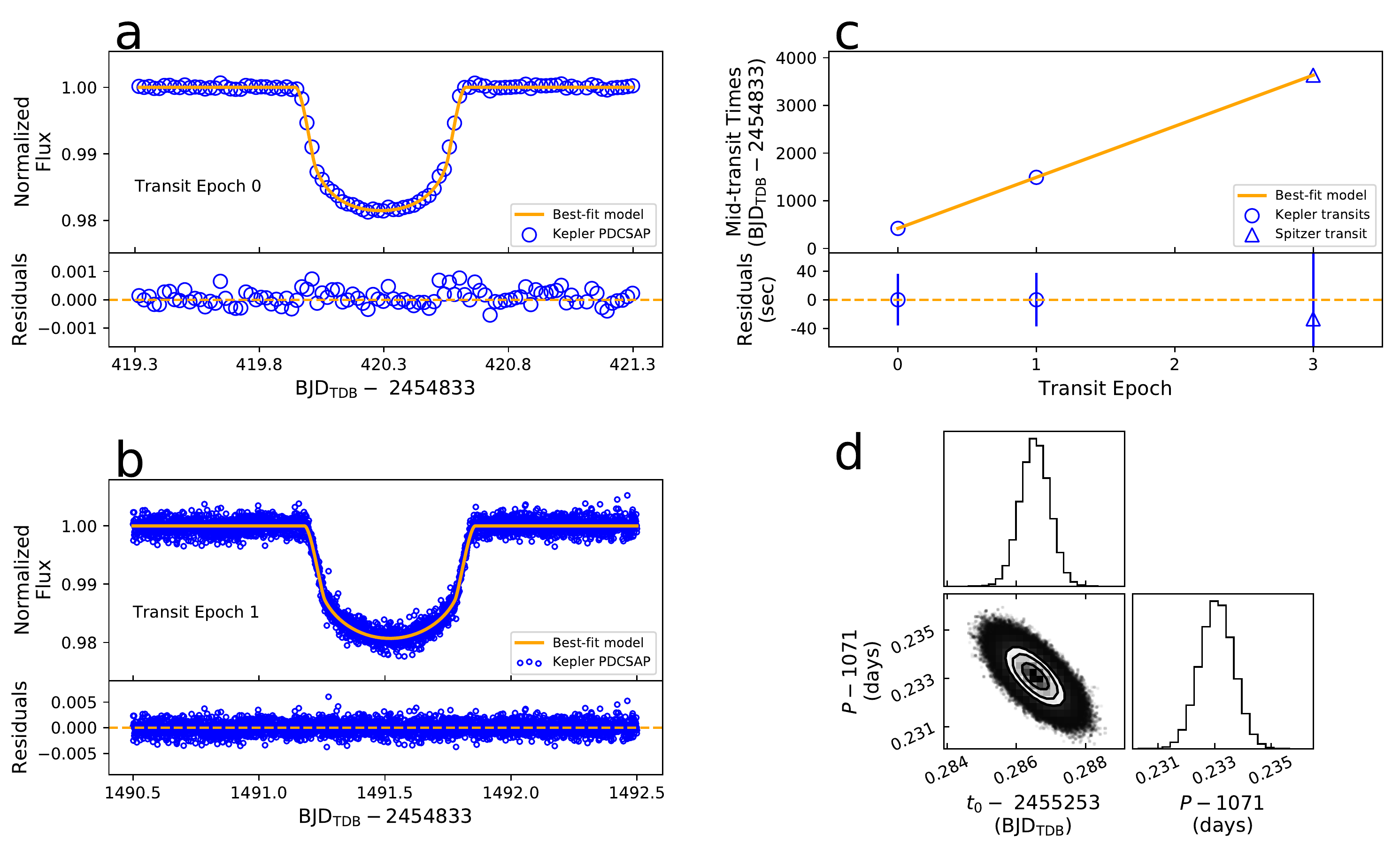}
    \caption{\Kepler\ light curves and best-fit models from Quarter 4 (Panel (a)) and Quarter 16 (Panel (b)). Panel (c): the three mid-transit times are well described by a linear model. The uncertainty in the \Spitzer\ transit time (epoch 3) is 23~minutes. Panel (d): posterior probability distributions for the parameters in the linear fit to the mid-transit times \citep[made in-part with \texttt{corner},][]{ForemanMackey2016a}.}
    \label{fig:transits_ephem}
\end{figure*}

TTVs can be effectively detected or ruled out for an exoplanet with at least three observed transits. Here, we measure individual mid-transit times of Kepler-167e in the \Kepler\ data and combine these with the \Spitzer\ transit time to quantify the extent to which TTVs affect the ephemeris. We acquired the pre-search data conditioning photometry (PDCSAP) from Data Release 25 for Quarters 4 (long cadence) and 16 (short cadence).\footnote{See \url{https://archive.stsci.edu/kepler/}. Accessed 2018 December 5.} These PDCSAP fluxes were processed with version 9.3 of the Science Operations Center pipeline, which included improvements in the calculation of the stellar scene that resulted in more accurate crowding metrics \citep{Dalba2017a} and higher photometric precision on average \citep[e.g.,][]{Twicken2016}. Near each transit, the PDCSAP flux was relatively free from low-frequency noise and other systematic errors (Figure \ref{fig:transits_ephem}, left panels), so we did not complete any additional processing. 

Both Kepler-167e transits were individually fit to transit models produced by the BAsic Transit Model cAlculatioN \citep[\texttt{BATMAN},][]{Kreidberg2015} package using the MCMC ensemble sampler \texttt{emcee} \citep{ForemanMackey2013}. Flux ``smearing'' caused by the 29.4~minute sampling of the long cadence data was accounted for through numerical integration \citep{Kipping2010,Southworth2011}.

For each transit, Gaussian priors were imposed on $P$, $a/R_{\star}$, $i$, and $e$ using the values reported by \citet{Kipping2016b}. Uniform priors spanning the entire physically allowed parameter space were imposed on $t_{\rm mid}$, $R_p/R_{\star}$, $\omega$, and the quadratic limb darkening parameters---sampled following \citet{Kipping2013b}. While the application of priors from \citet{Kipping2016b} was somewhat circular (i.e., choosing priors based on analysis of a data set containing the same transit being fit), it assured that parameters that could not be precisely inferred from a single transit were constrained by the other transit. We gauged convergence of the fit by comparing the integrated autocorrelation time ($\tau$) to the length of the chains \citep[e.g.,][]{ForemanMackey2013}. Sampling continued until the chain lengths for all fitted parameters were at least 10$\tau$ (50$\tau$ in the case of $t_{\rm mid}$). For both transits, the effective number of samples of the $t_{\rm mid}$ posterior was of order 1.5\e{4}.  The results of the \Kepler\ fits are presented in Table \ref{tab:transits} and Figure \ref{fig:transits_ephem} (left panels). 

\begin{deluxetable}{cccc}
    \tablecaption{Timing of Past and Future Transits of Kepler-167e As Calculated in This Work} 
    \tablecolumns{4}
    \tablewidth{0pt}
    \tablehead{ 
        & 
        \colhead{$t_{\rm mid}$} &
        \colhead{$t_{\rm mid}$} &
        \colhead{$\sigma_{t_{\rm mid}}$\tablenotemark{a}}  \\
        \colhead{Epoch} &
        \colhead{(BJD$_{\rm TDB}$)} &
        \colhead{(UTC)} &
        \colhead{(minutes)}
        }
    \startdata 
        0 & 2455253.28654 & 2010 Feb 25 18:51:27 & $\pm$0.60 \\
        1 & 2456324.51962 & 2013 Feb 1 00:27:08 & $\pm$0.62 \\
        2\tablenotemark{b} & \nodata & \nodata & \nodata \\
        3 & 2458466.985 & 2018 Dec 14 11:38 & $\pm$23 \\
        4\tablenotemark{c} & 2459538.2189 & 2021 Nov 19 17:14 & $\pm$3.0 \\
        5 & 2460609.4519 & 2024 Oct 25 22:50 & $\pm$3.9 \\
        6 & 2461680.6850 & 2027 Oct 2 04:25 & $\pm$4.7 \\
        7 & 2462751.9181 & 2030 Sep 7 10:01 & $\pm$5.6
    \enddata
    \tablenotetext{a}{Mid-transit time uncertainties ($\sigma_{t_{\rm mid}}$) are measured or predicted for transits in the past or future, respectively.}
    \tablenotetext{b}{This transit occurred after the primary \Kepler\ mission ended.}
    \tablenotetext{c}{This transit is not observable by \JWST\ due to solar avoidance constraints (see Section \ref{sec:future}).}
    \label{tab:transits}
\end{deluxetable} 

Kepler-167e's three mid-transit times were fit with a linear model as a function of epoch (Figure \ref{fig:transits_ephem}, top right panel). The sampler \texttt{emcee} was again used to infer the fitted parameters and their uncertainties. Convergence of the fit was judged in the same fashion as before. The posterior probability distributions of the two linear parameters are shown in Figure \ref{fig:transits_ephem} (bottom right panel). 

For Kepler-167e, we report an orbital period of $P~=~1071.23308^{+0.00059}_{-0.00060}$~days and an initial transit epoch of $t_0~=~2455253.28654^{+0.00042}_{-0.00042}$~BJD$_{\rm TDB}$. These values are consistent with the ephemeris from \citet{Kipping2016b}. Therefore, the \Spitzer\ observations have confirmed the accuracy of the linear ephemeris inferred from the two \Kepler\ transits and enabled the accurate prediction of future transit times. The magnitude of the uncertainty on $P$ calculated in this work and by \citet{Kipping2016b} is similar because of the relatively large uncertainty on the timing of the \Spitzer\ transit compared to the \Kepler\ transits. However, now the reported orbital period of Kepler-167e does not contain an unaccounted for source of uncertainty due to the unknown influence of TTVs. 

We place upper limits on TTVs in the ephemeris of Kepler-167e by comparing the time intervals between subsequent transits. Epochs 0 and 1 are separated by $1071.23308\pm0.00060$~days, and epochs 1 and 3 are separated by $2142.465\pm0.016$~days. Dividing the latter in half, we find the difference between these two orbital period estimates to be $0.00058\pm0.0079$~days, meaning that we do not measure a variation in the timing of the Kepler-167e transits. Based on our measurement uncertainty ($\sigma$), we rule out TTVs of order 11, 34, and 57~minutes to $1\sigma$, $3\sigma$, and $5\sigma$ significance, respectively.    

The extent to which TTVs are ruled out is largely determined by the precision of the \Spitzer\ observations. As mentioned in Section \ref{sec:spitzer_obs}, the choice to observe in Channel 1 involved a direct trade-off between photometric precision and the test for hydrocarbon opacity near 3.6~$\mu$m. On the other hand, from only a partial transit, neither Channel would have been capable of matching the transit timing precision of the \Kepler\ data, which indeed dominates the final ephemeris.

\subsection{Future Kepler-167e Transit Times}\label{sec:future}

Using the posterior probability distributions of the parameters in Kepler-167e's linear transit ephemeris, we predicted its mid-transit times through 2030 (Table \ref{tab:transits}). In all cases, the uncertainty is less than six minutes, making Kepler-167e an accessible target to telescopes on which time is expensive and limited.

We used the \HST\ Astronomer's Proposal Tool\footnote{See \url{http://www.stsci.edu/hst/proposing/apt}.} to determine that all transits through 2030 are observable by \HST. We then used the \JWST\ General Target Visibility Tool\footnote{See \url{https://jwst-docs.stsci.edu/display/JPP/JWST+General+Target+Visibility+Tool+Help}.} to determine that Kepler-167's annual window of \JWST\ observability (based on solar elongation) spans mid-April to early November. In 2021, this window closes on approximately November 8. As a result, the 2021 transit of Kepler-167e cannot be observed by \JWST. All other transits of Kepler-167e through 2030 will be observable by \JWST.


\section{Discussion: The Rarity of a Transiting Jupiter-analog}\label{sec:disc}

The transiting nature of Kepler-167e makes it a remarkably improbable target for follow-up observation. It is a member of a small group of transiting exoplanets that orbit their host stars from beyond the water ice line. Moreover, the timing of Kepler-167e's future transits are not plagued by uncertainties from poorly constrained TTVs \citep{Wang2015}. Based on the \Spitzer\ observations presented here, follow-up efforts for Kepler-167e are no longer high risk, only high-reward. 

Kepler-167e bares several striking resemblances to Jupiter and Saturn that make it an exceptional target for future study. Its radius is within 10\% of Jupiter's and Saturn's. Its equilibrium temperature ($T_{\rm eq}~\approx~131$~K, assuming a Jupiter-like Bond albedo) is only 20~K warmer than Jupiter's. Its eccentricity ($e~=~0.062^{+0.104}_{-0.043}$) is consistent with Jupiter's ($e~=~0.0489$) and Saturn's ($e~=~0.0565$). These properties highlight potential connections between these planets' formation and evolution histories that warrant further investigation.


\subsection{The Next Steps in the Characterization of Kepler-167e}

The immediate next step in the characterization of Kepler-167e is the measurement of its mass ($M_p$). Our references to Kepler-167e as an exoplanet have made the assumption that its mass is planetary. \citet{Kipping2016b} argued that Kepler-167e is four times more likely to be planetary than sub-stellar, but its mass has not yet been measured. Obtaining the RV precision to distinguish between these two scenarios is likely possible from current precision-RV facilities assuming a sufficiently long baseline of observations is collected.  

After Kepler-167e's mass is measured, the next step is the characterization of its atmosphere via transmission spectroscopy. Using the \Spitzer\ $R_p/R_{\star}$ measurement, we approximate the 3.6~$\mu$m transit depth of Kepler-167e as $(R_p/R_{\star})^2_{\rm Spitzer}~=~1.589\pm0.088$\%. This is indistinguishable from the more precise \Kepler\ transit depth: $(R_p/R_{\star})^2_{\rm Kepler}~=~1.641\pm0.017$\% \citep{Kipping2016b}. Featureless (flat) transmission spectra have previously been interpreted as signs of clouds or high mean molecular weight atmospheres \citep[e.g.,][]{Kreidberg2014a}. However, we caution that more data are needed before an interpretation of Kepler-167e's atmosphere can be made.  

\begin{figure}
    \centering
    \includegraphics[width=\columnwidth]{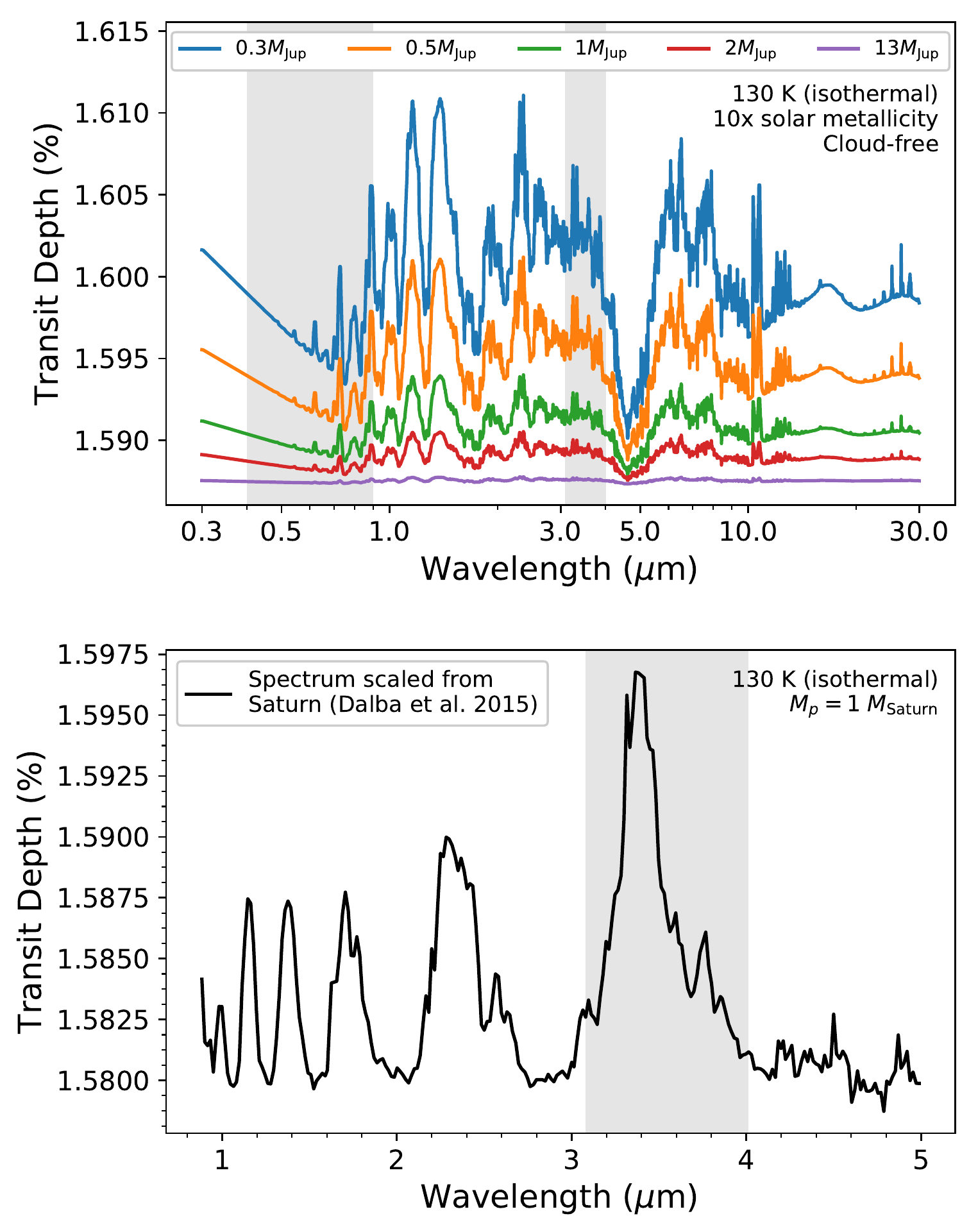}
    \caption{Models of potential transmission spectra of Kepler-167e. Gray regions denote \Kepler\ and \Spitzer-IRAC Channel 1 bandpasses. Top: spectra modeled using \texttt{Exo-Transmit}. The 13~$M_{\rm Jupiter}$ spectrum is consistent with the \Spitzer\ transit depth and the others are offset for clarity. The CH$_4$ absorption between 1 and 2~$\mu$m and the deep window at 5~$\mu$m points to the potential benefit of conducting joint \HST-\JWST\ transmission spectroscopy. Bottom: spectrum based on Saturn \citep{Dalba2015} but scaled to the properties of Kepler-167e. The 3.4~$\mu$m feature is partially due to opacity from CH$_4$ photochemical byproducts. The potential to detect an active methanological in Kepler-167e's atmosphere motivates additional modeling and future near-infrared transmission spectroscopy.}
    \label{fig:tts}
\end{figure}

In Figure \ref{fig:tts}, we display several transmission spectrum models that may describe Kepler-167e. Here, we discuss these models and their implications for future observations. 

First, we utilize \texttt{Exo-Transmit} \citep{Kempton2017,Freedman2008,Freedman2014,Lupu2014} to model transmission spectra of Kepler-167e for several masses assuming a 130~K isothermal, cloud-free atmosphere. We apply an equation of state that assumes equilibrium chemistry at 10 times the solar metallicity that includes the condensation of molecular species. The chosen masses span the planetary regime of the 16th--84th percentile uncertainty range predicted by \texttt{forecaster} \citep{Chen2017}. At 130~K, many of the features are indicative of CH$_4$ absorption. For H$_2$O, deeper layers of the atmosphere act as a cold trap, condensing it into a cloud layer. Similarly, NH$_3$ likely condenses (as is the case on Jupiter and Saturn). As expected, the size of the features scales strongly with $M_p$. For the $0.3~M_{\rm Jupiter}$ scenario, the peaks of the 1--2~$\mu$m CH$_4$ features are $\sim$200~parts per million (ppm) above the deep 5~$\mu$m window, indicating the potential amenability of Kepler-167e to joint \HST-\JWST\ transmission spectroscopy. Even at $M_p=1~M_{\rm Jupiter}$, this peak-to-trough difference is $\sim$50~ppm.

Cloud layers and refraction boundaries \citep[e.g.,][]{Misra2014a} have not been included in the \texttt{Exo-Transmit} analysis. Clouds and refraction can truncate transmission spectra features at particular pressure levels thereby reducing their contrast. These effects can lead to degenerate interpretations of the spectra. Furthermore, \texttt{Exo-Transmit} does not include ice condensation or molecular diffusion \citep{Kempton2017}, which may occur in cold atmospheres. These processes can alter the chemical abundances and structure of an atmosphere and ultimately may change the nature and size of transmission spectrum features.

Second, we draw analogy to the solar system by scaling the reconstructed transmission spectrum of Saturn \citep{Dalba2015} to Kepler-167e assuming $M_p~=~1~M_{\rm Saturn}$ (Figure \ref{fig:tts}, bottom panel). This spectrum is shifted to be consistent with the measured \Spitzer\ transit depth. This comparison to Saturn enables the consideration of many of the complications that are not included in the \texttt{Exo-Transmit} models (e.g., photochemistry, ice condensation, and atmospheric refraction). The 50--75~ppm features blueward of 3~$\mu$m are caused by CH$_4$ absorption, while the largest feature near 3.4~$\mu$m indicates additional opacity from larger hydrocarbons produced via CH$_4$ photolysis \citep{Dalba2015}. If similar hydrocarbon photochemistry is ongoing in the atmosphere of Kepler-167e, it can potentially be identified and characterized via near-infrared transmission spectroscopy with \JWST. 

The baseline of Kepler-167e's Saturn-based transmission spectrum is set by atmospheric refraction \citep{Dalba2015}. The importance of refraction in exoplanet transit observations can be assessed quantitatively \citep[e.g.,][]{Hui2002,Dalba2017b}. Based on the criterion of \citet{Dalba2017b}, there is a region of parameter space (spanning planet masses, atmospheric properties, and observation wavelength) for which refraction would set the baseline of Kepler-167e's transmission spectrum. For this region, the out-of-transit flux increase caused by refraction \citep[e.g.,][]{Sidis2010,Dalba2017b} would also be present. Since transmission spectroscopy typically requires the observation of pre- and post-transit baseline, the out-of-transit flux increase would provide additional atmospheric information for little or no addition in observing time. However, we caution that an informed prediction of the magnitudes of refraction effects in future observations of Kepler-167e requires a mass measurement and additional modeling.


\subsection{Concluding Remarks}

The \Spitzer\ observations presented here enable the accurate prediction of future transit times of Kepler-167e with a precision better than six minutes. This makes Kepler-167e a strong candidate for transmission spectroscopy that is accessible to \HST\ and \JWST. Although we have not simulated observations from these observatories---mainly due to the uncertainty introduced by Kepler-167e's presently unknown mass---the integration of signal over its 16 hr long transit will likely enable novel investigations of cold exoplanet atmospheres. Kepler-167e's many similarities to Jupiter place it in a truly rare category of transiting exoplanets and identify it as a benchmark for future comparative planetology investigations.


\acknowledgments

The authors thank the anonymous referee for an insightful review that improved the quality of this work.
This work is based on observations made with the \textit{Spitzer Space Telescope}, which is operated by the Jet Propulsion Laboratory, California Institute of Technology, under a contract with NASA. The authors acknowledge Philip Muirhead for contributing to the scientific development of this research. The authors also acknowledge Stephen Kane and Heather Knutson for helpful conversations and Drake Deming for providing a code to analyze full-frame \Spitzer\ observations. Some of the data presented in this paper were obtained from the Mikulski Archive for Space Telescopes (MAST). STScI is operated by the Association of Universities for Research in Astronomy, Inc., under NASA contract NAS5-26555.

\vspace{5mm}
\facility{\Spitzer, \Kepler} 

\vspace{5mm}
\software{\\    \texttt{astropy} \citep{astropy2013},\\
                \texttt{BATMAN} \citep{Kreidberg2015}, \\
                \texttt{corner} \citep{ForemanMackey2016a},\\
                \texttt{emcee} \citep{ForemanMackey2013}, \\
                \texttt{Exo-Transmit} \citep{Kempton2017}, \\
                \texttt{forecaster} \citep{Chen2017}
                }

\end{document}